# Unruh effect as a noisy quantum channel

Doyeol Ahn[1,2*]

[1]*Department of Electrical and Computer Engineering and Center for Quantum Information Processing, University of Seoul, Seoul 02504, Republic of Korea*

[2]*Physics Department, Charles E. Schmidt College of Science, Florida Atlantic University, 777 Glades Road, Boca Raton, FL 33431-0991, USA*

*Abstract:* We studied the change of the nonlocal correlation of the entanglement in Rindler spacetime by showing that the Unruh effect can be interpreted as a noisy quantum channel having a complete positive and trace preserving map with an "operator sum representation." It is shown that the entanglement fidelity is obtained in analytic form from the "operator sum representation", which agrees well numerically with the entanglement monotone and the entanglement measure obtained previously. Non-zero entropy exchange between the system $Q$ and the region $II$ of the Rindler wedge indicates the nonlocal correlation between casually disconnected regions. We have also shown the sub additivity of entropies numerically.

*To whom correspondence should be addressed.
E-mail: dahn@uos.ac.kr: daveahn@ymail.com

The Unruh effect [1-4] discovered over four decades ago predicts that a non-inertial observer in an accelerated motion would see the Minkowski vacuum as thermal bath of excited particles. The discovery of Unruh effect is regarded as one of the monumental achievements of our understanding of quantum field theory in curved space-time despite the lack of direct experimental confirmation. Recently, there has been renewed interest of the Unruh effect especially associated with the entanglement harvesting [5-8] and the detection of possible signature of Unruh effect in the quantum radiation [9-11]. Iso et al. [11] pointed out that this quantum radiation is related to the nonlocal correlation nature of the Minkowski vacuum state, which has its origin in the entanglement of the state between the left and the right Rindler wedges.

In this work, we study the change of the nonlocal correlation of the entanglement in Rindler space from a quantum information point of view by showing that the Unruh effect can be interpreted as a noisy quantum channel having a complete positive and trace preserving map with an "operator sum representation." The setting, in which Alice and Rob are two observers, one inertial and the other non-inertial, describes the entanglement between two modes of free scalar field from the point of their detectors [12-14]. When a non-inertial observer, Rob, is under the influence of the acceleration, the measure of entanglement seen by the non-inertial observer is affected by the presence of quantum thermal fields. The state observed by an inertial observer Alice and a non-inertial observer Rob is an $2\times\infty$ dimensional space in which case the necessary and sufficient criteria for the entanglement is not so well established [14]. When a quantum system is coupled to the Unruh radiation, it is inevitably treated in an infinite dimensional space, in which case only a Gaussian state has an entanglement measure [15-17]. For this reason, Alsing and Milburn [12] used an indirect measure of entanglement as they calculated teleportation fidelity. Fuentes-Schuller and Mann [13] calculated lower bound of entanglement. Ahn and Kim [14] studied an entanglement measure by calculating the symplectic eigenvalues of the matrix obtained through the partial transposition of the variance matrix.

Here, we obtain the entanglement fidelity directly from the "operator sum representation [18]" of the complete positive super-operator $\varepsilon^Q$, which acts on the initial density operator $\rho^Q$ in analytical form. It is shown that our analytical result agrees very well with the entanglement monotone [13] and the entanglement measure [14] obtained numerically. We assume that the quantum state $Q$ describes an entanglement between Alice and Rob in stationary states, i.e., the state in which Rob also stays stationary without acceleration. We will describe the evolution of the system $Q$ by allowing Rob to experience uniform acceleration $a$ through the acceleration parameter $r$ defined by $\tanh r = \exp(-2\pi\Omega)$, $\Omega = |k|c/a$, $k$ the wave vector, $c$ the speed of light, $a$ the uniform acceleration. We consider the real, scalar field both in the Minkowski and the Rindler spacetime. Let Alice be an observer at event $P$ with zero velocity in the Minkowski spacetime and non-inertial observer Rob be moving with positive uniform acceleration in the z direction with respect to Alice (Fig. 1). If Rob is under a uniform acceleration, the corresponding ground state should be specified in Rindler coordinate [19-21] in order to describe what Rob observes. Let us denote the ground states, which Alice and Rob detect in the Minkowski spacetime as $|O_A\rangle_M$ and $|O_R\rangle_M$ (Fig. 1), respectively. Then ground state from the non-inertial point of view can be written as $|O_R\rangle_M = \frac{1}{\cosh r}\sum_{n=0}^{\infty}\tanh^n r |n\rangle_I \otimes |n\rangle_{II}$, with $|n\rangle_I$ and $|n\rangle_{II}$ the mode decompositions in Rindler regions $I$ and $II$, respectively [14]. The excited state for Rob in Minkowski spacetime is obtained by applying the Minkowski creation operator $a_R^\dagger$ to the vacuum state successively [14]. For example,

$$|1_R\rangle_M = a_R^\dagger |O_R\rangle_M, \quad |2_R\rangle_M = \tfrac{1}{\sqrt{2!}}(a_R^\dagger)^2 |O_R\rangle_M, \quad ... \quad |m_R\rangle_M = \frac{1}{\sqrt{m!}}(a_R^\dagger)^m |O_R\rangle_M. \qquad (1)$$

The particle creation and annihilation operators for the Rindler space-time are expressed as $b_\sigma^\dagger$ and $b_\sigma$, respectively. Here, the subscript $\sigma = I$ or $II$, takes into account the fact that the space-time has an event horizon, so that it is divided into two causally disconnected Rindler wedges $I$ and $II$ (Fig. 1). The Minkowski operators $a_R^\dagger$ and $a_R$ can be expressed in terms of the Rindler operators $b_\sigma^\dagger$ and $b_\sigma$ by Bogoliubov transformations [14]:

$$a_R^\dagger = b_I^\dagger \cosh r - b_{II} \sinh r = G b_I^\dagger G^\dagger, \quad a_R = b_I \cosh r - b_{II}^\dagger \sinh r = G b_I G^\dagger, \qquad (2)$$

with $G = \exp\{r(b_I^\dagger b_{II}^\dagger - b_I b_{II})\}$. Then, the Minkowski ground state $|O_R\rangle_M$ seen by the Rindler observer, i.e., Rob, is given by $|O_R\rangle_M = G(|O\rangle_I \otimes |O\rangle_{II})$. This is the basis of the Unruh effect, which says that a non-inertial observer with uniform acceleration

would see thermal quantum fields. In other words, Rob would see the quantum bath populated by thermally excited states. The excited states for Rob in Minkowski spacetime are now given by [14]

$$a_R^\dagger |O_R\rangle_M = Gb_I^\dagger(|O\rangle_I \otimes |O\rangle_{II}), ..., (a_R^\dagger)^m |O_R\rangle_M = G(b_I^\dagger)^m(|O\rangle_I \otimes |O\rangle_{II}). \quad (3)$$

We now consider the system $Q'$ described by

$$\rho^{Q'} = Tr_{II}(|\psi\rangle\langle\psi|), \quad (4)$$

where $|\psi\rangle = \frac{1}{\sqrt{2}}(|0_A\rangle_M \otimes |1_R\rangle_M + |1_A\rangle_M \otimes |0_R\rangle_M) \quad (5)$

and $Tr_{II}$ denotes the partial trace over the states of Rindler wedge II. The initial quantum state $\rho^Q$ is given by

$$\rho^Q = \lim_{a\to 0}\left[ Tr_{II}(|\psi\rangle\langle\psi|)\right]$$
$$= \frac{1}{2}(|01\rangle + |10\rangle)(\langle 01| + \langle 10|), \quad (6)$$

where $|nm\rangle = |n_A\rangle \otimes |m_R\rangle_I$.

Here, we would like to treat the Unruh effect as a noisy quantum channel [18] where the system $Q$ prepared in an initial state $\rho^Q$ is described by the dynamical process, after which the system is in $\rho^{Q'}$. The dynamical process is described by a map $\varepsilon^Q$, so that the evolution is [18]

$$\rho^Q \to \rho^{Q'} = \varepsilon^Q(\rho^Q). \quad (7)$$

If the map $\varepsilon^Q$ is given by

$$\varepsilon^Q(\rho^Q) = \sum_n A_n^Q \rho^Q A_n^{Q\dagger} \quad (8)$$

where $A_n^Q$ is an operator on the Hilbert space of $Q$ only, then the map is a completely positive map [18]. From Eqs. (1) to (5), we obtain after some mathematical manipulation [13]

$$\rho^{Q'} = \frac{1}{\cosh^2 r}\sum_{n=0}^{\infty}(\tanh^2 r)^n \rho_n = \rho_{AR} \quad (9)$$

with
$$\rho_n = \frac{1}{2}\left\{|1n\rangle\langle 1n| + \frac{\sqrt{n+1}}{\cosh r}(|1n\rangle\langle 0(n+1)| + |0(n+1)\rangle\langle 1n|) \right.$$
$$\left. + \frac{(n+1)}{\cosh^2 r}|0(n+1)\rangle\langle 0(n+1)|\right\}. \quad (10)$$

By comparing Eqs. (6) and (10), we obtain

$$A_n^Q = \frac{1}{\sqrt{n!}} \frac{\tanh^n r}{\cosh^2 r} (\cosh r)^{\hat{n}_A} \otimes (b_I^\dagger)^n \qquad (11)$$

where $\hat{n}_A = a_A^\dagger a_A$ is a number operator acting on Alice's Hilbert space. From this one can see that the Unruh effect can be described by a completely positive map acting on the quantum state $Q$ of Alice and Rob, when both parties are in the stationary state, i.e., zero acceleration for Rob. Let $|\phi^Q\rangle$ be a quantum state of $\rho^Q$, then after some manipulation we obtain

$$\begin{aligned}
&\sum_{n=0}^\infty \langle \phi^Q | A_n^{Q\dagger} A_n^Q | \phi^Q \rangle \\
&= \frac{1}{2} \sum_{n=0}^\infty \frac{(\tanh^2 r)^n}{\cosh^2 r} \left( \langle 1n| + \frac{\sqrt{n+1}}{\cosh r} \langle 0(n+1)| \right) \left( |1n\rangle + \frac{\sqrt{n+1}}{\cosh r} |0(n+1)\rangle \right) \\
&= \frac{1}{2\cosh^2 r} \sum_{n=0}^\infty (\tanh^2 r)^n \left( 1 + \frac{n+1}{\cosh^2 r} \right) \\
&= 1 \\
&= Tr\left( \sum_{n=0}^\infty A_n^Q |\phi^Q\rangle \langle \phi^Q | A_n^{Q\dagger} \right) \\
&= Tr \rho^{Q'}.
\end{aligned} \qquad (12)$$

This indicates that the map is trace preserving. The map is complete positive, trace preserving and as a result can be represented by an "operator sum representation" [18]. The Unruh effect transforms the stationary entangled state into the mixed state in Rindler space by a complete positive trace preserving map. Here, we have used the following relations:

$$\begin{aligned}
&\frac{1}{\cosh^2 r} \sum_{n=0}^\infty (\tanh^2 r)^n \\
&= \frac{1}{\cosh^2 r} \frac{1}{(1-\tanh^2 r)} \\
&= \frac{1}{\cosh^2 r} \frac{\cosh^2 r}{(\cosh^2 r - \sinh^2 r)} \\
&= 1
\end{aligned} \qquad (13)$$

and

$$\frac{1}{\cosh^4 r}\sum_{n=0}^{\infty}(n+1)\left(\tanh^2 r\right)^n$$

$$=\frac{1}{\cosh^4 r}\frac{d}{d\left(\tanh^2 r\right)}\sum_{n=0}^{\infty}\left(\tanh^2 r\right)^{n+1}$$

$$=\frac{1}{\cosh^4 r}\frac{d}{d\left(\tanh^2 r\right)}\frac{\tanh^2 r}{1-\tanh^2 r} \quad (14)$$

$$=\frac{1}{\cosh^4 r}\frac{1}{\left(1-\tanh^2 r\right)^2}$$

$$=1.$$

According to Schumacher [18], for complete positive and trace preserving map, the entanglement fidelity $F_e$ which measures how successfully the quantum channel preserves the entanglement of $Q$ can be represented by

$$F_e = \sum_n \left(Tr\rho^Q A_n^Q\right)\left(Tr\rho^Q A_n^{Q\dagger}\right). \quad (15)$$

From Eqs. (6) and (11), we obtain

$$Tr\rho^Q A_n^Q$$
$$= Tr\left[\frac{1}{2}(|10\rangle+|01\rangle)(\langle 10|+\langle 01|)\frac{1}{\sqrt{n!}}\frac{\tanh^n r}{\cosh^2 r}(\cosh r)^{\hat{n}_A}\otimes\left(b_I^\dagger\right)^n\right] \quad (16)$$
$$= \frac{1}{2}\frac{1}{\cosh r}\left(1+\frac{1}{\cosh r}\right)\delta_{n,0}$$

and as a result

$$F_e = \sum_n \left(Tr\rho^Q A_n^Q\right)\left(Tr\rho^Q A_n^{Q\dagger}\right)$$
$$= \frac{1}{4}\frac{1}{\cosh^2 r}\left(1+\frac{1}{\cosh r}\right)^2. \quad (17)$$

When Rob is in stationary state $a\to 0$ and $\cosh r \to 1$. Then, from Eq. (17) the entanglement fidelity approaches unity, i.e., $F_e \to 1$. On the other hand, when the value of the acceleration is large, then $\cosh r$ is increasing monotonically and the entanglement fidelity also decreases monotonically approaching zero for very large value of the acceleration (Fig. 2). Our analytical result for the entanglement fidelity agrees very well with the entanglement monotone obtained by Feuntes-Schuller and Mann [13] and Ahn and Kim [14].

Since the final state $|\psi\rangle$ is a pure state, the von Neumann entropy $S(|\psi\rangle\langle\psi|)=0$ and as a result, we obtain

$$S(\rho^{Q'}) = S(\rho_{AR}) = S(\rho_{II}) \tag{18}$$

where $\rho_{II} = Tr_{AI}(|\psi\rangle\langle\psi|)$. The entropy defined by Eq. (18) is called an entropy exchange $S_e$ [18], which is common entropy for two initially uncorrelated systems. Another measure of correlation is the mutual information $I(\rho_{AR})$, which is defined by [13]

$$I(\rho_{AR}) = S(\rho_A) + S(\rho_R) - S(\rho_{AR}) \tag{19}$$

where $\rho_A = Tr_I(\rho_{AR})$ and $\rho_R = Tr_A(\rho_{AR})$. The detailed expressions for entropies are given by [13]

$$S(\rho_{AR}) = -\sum_n a_n \left(1 + \frac{n+1}{\cosh^2 r}\right) \log_2 \left[a_n \left(1 + \frac{n+1}{\cosh^2 r}\right)\right], \tag{20}$$

$$S(\rho_R) = -\sum_n a_n \left(1 + \frac{n}{\sinh^2 r}\right) \log_2 \left[a_n \left(1 + \frac{n}{\sinh^2 r}\right)\right], \tag{21}$$

$$S(\rho_A) = 1, \tag{22}$$

with $\quad a_n = \frac{(\tanh^2 r)^n}{2\cosh^2 r}$. $\tag{23}$

The mutual information $I(\rho_{AR})$ is a measure of total correlation between Alice and Rob in the their entangled state. In Fig. 3 we plot the entropy exchange (solid line) and the mutual information (dashed line) as a function of the acceleration r. As acceleration increases the mutual information is approaching unity, which indicates that the states become mixed state [13]. From Eq. (6), the eigenvalues of the reduced density matrix $\rho_{AR}$ for $r \to 0$ are 0,0,0,1 and as a result we have $S(\rho_{AR}) = 0$. On the other hand, when the acceleration becomes infinite, we have $a_n \left(1 + \frac{n+1}{\cosh^2 r}\right) \to 0$ and as a result

$$a_n \left(1 + \frac{n+1}{\cosh^2 r}\right) \log_2 a_n \left(1 + \frac{n+1}{\cosh^2 r}\right) \to 0 \tag{24}$$

and we obtain $S(\rho_{AR}) \to 0$. The peak value of the entropy exchange exceeds 2 and this is the amount of correlation that Alice and Rob's entangled states have with the quantum bath due to the Unruh effect. In Fig. 4, we show the sub-additivity [18] $S_e = S(\rho_{AR}) \leq S(\rho_A) + S(\rho_R)$ numerically. According to the interpretation of non-relativistic quantum information theory [18], the entropy exchange characterizes the information exchange between the system $Q$ and the external world during the evolution given by $\mathcal{E}^Q$. Since, the region I and region II of the Rindler wedges are causally disconnected and the entropy exchange as the information exchange between the system $Q$ and causally disconnected external world, i.e., region II of Rindler wedge, can be interpreted as a measure of non-local correlation.

In summary, we studied the change of the nonlocal correlation of the entanglement in Rindler spacetime by showing that the Unruh effect can be interpreted as a noisy quantum channel having a complete positive and trace preserving map with an "operator sum representation." It is shown that the entanglement fidelity is obtained in analytic form, which agrees well with entanglement monotone [13] and the entanglement measure [14], numerically. Non-zero entropy exchange between the system $Q$ and the region *II* of the Rindler wedge indicates the nonlocal correlation between casually disconnected regions. We have also shown sub additivity of entropies numerically.

**Acknowledgements**

This work was supported by Institute for Information & communications Technology Promotion(IITP) grant funded by the Korea government(MSIP) and the US Air Force (BAA-AFRL-AFOSR-2016-0007) (No.2017-0-00266, Gravitational effects on the free space quantum key distribution for satellite communication).

**Figure captions**

**Figure 1** Rindler spacetime. In region *I* and *II*, time coordinates $\eta =$ constant are straight lines through the origin. Space coordinates $\zeta =$ constant are hyperbolae with null asymptotes $H_+$ and $H_-$, which act as event horizons. The Minkowski coordinates *t, z* and Rindler coordinates $\eta, \zeta$ are given by $t = a^{-1}\exp(a\zeta)\sinh a\eta$ and $z = a^{-1}\exp(a\zeta)\cosh a\eta$, where $a$ is a uniform acceleration (Reference 21). Alice and Rob initially share a two-mode squeezed state at the event *P*. We consider the case of Alice in stationary and Rob (green hyperbola) under uniform acceleration.

**Figure 2** Entanglement fidelity $F_e$ versus acceleration r. This measure of entanglement is obtained in analytical form, i.e., $F_e = \frac{1}{4\cosh^2 r}\left(1 + \frac{1}{\cosh^2 r}\right)^2$ as a function of the acceleration *r* from the "operator sum representation [18]." The results agree well with the entanglement monotone [13,14].

**Figure 3** Comparison of mutual information $I(\rho_{AR})$ and entropy exchange $S_e$. The mutual information is a measure of total correlation between Alice and Rob in the their entangled state while the entropy exchange is a common entropy for two initially uncorrelated systems.

**Figure 4** Numerical proof of sub additivity for the entropy $S(\rho_{AR}) \leq S(\rho_A) + S(\rho_R)$.

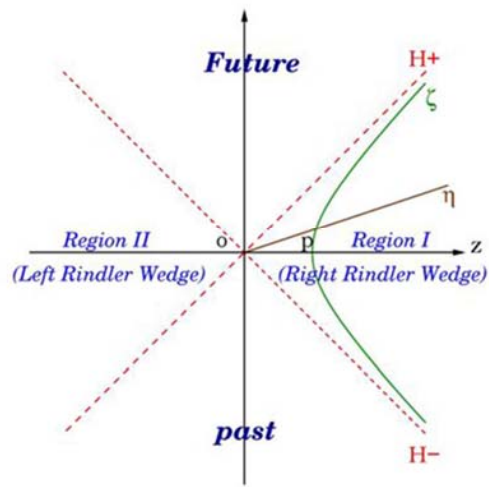

Fig. 1

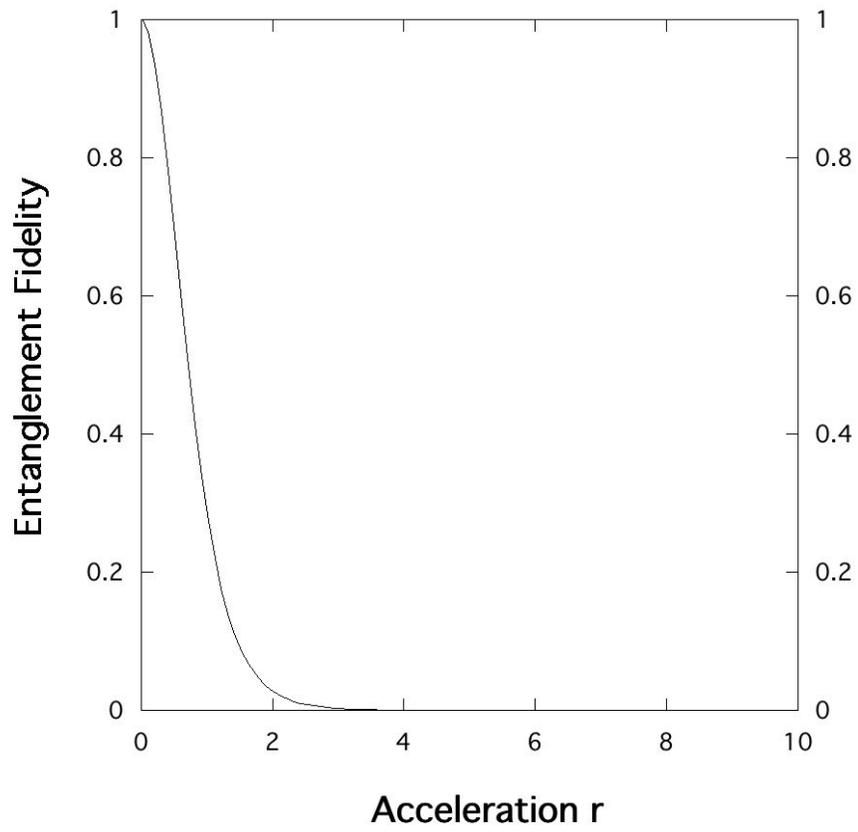

Fig. 2

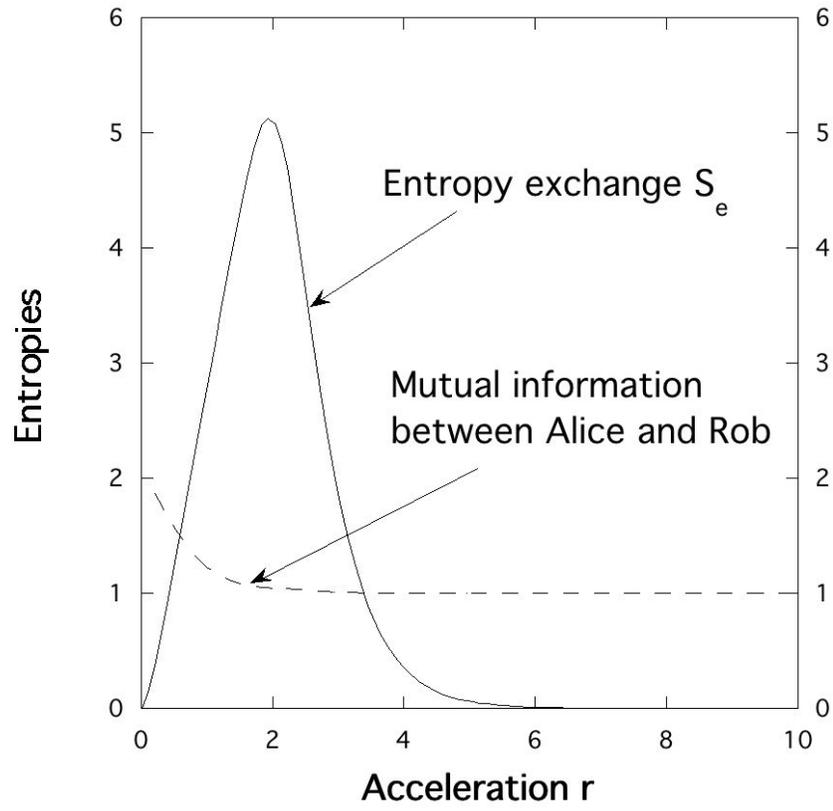

Fig. 3

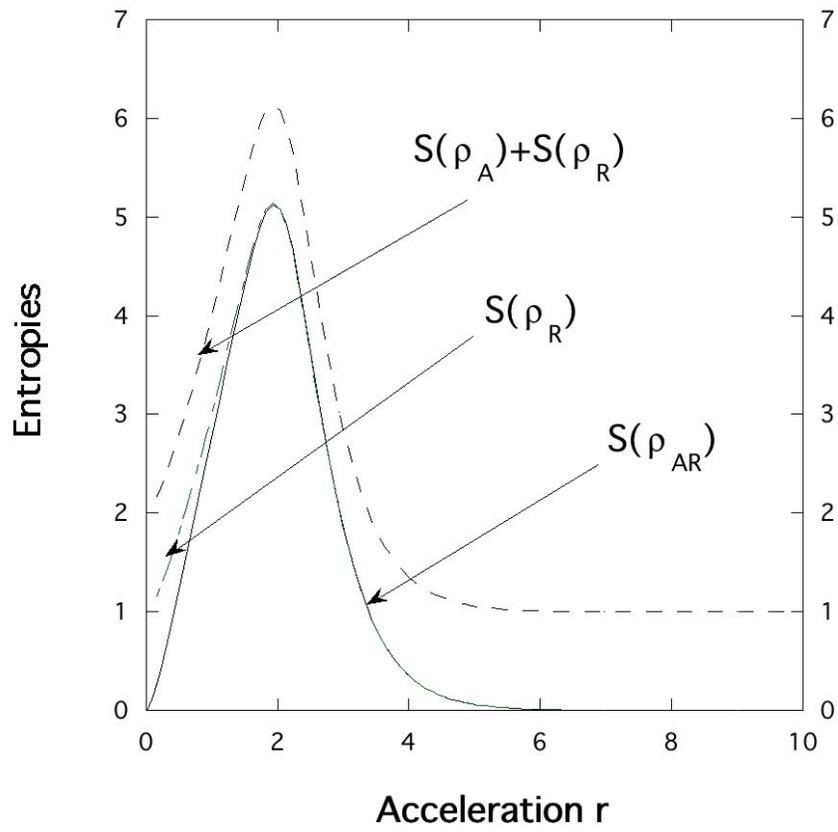

Fig. 4